White Paper for Heliophysics 2050

# Major Scientific Challenges and Opportunities in Understanding Magnetic Reconnection and Related Explosive Phenomena in Solar and Heliospheric Plasmas

H. Ji[1] and J. Karpen[2]

September 2020

Co-authors:

A. Alt[1], S. Antiochos[2], S. Baalrud[3], S. Bale[4], P. M. Bellan[5], M. Begelman[6], A. Beresnyak[7], A. Bhattacharjee[1], E.G. Blackman[8], D. Brennan[1], M. Brown[9], J. Buechner[10], J. Burch[11], P. Cassak[12], B. Chen[13], L.-J. Chen[2], Y. Chen[6], A. Chien[1], L. Comisso[14], D. Craig[15], J. Dahlin[16], W. Daughton[17], E. DeLuca[18], C. F. Dong[1], S. Dorfman[19], J. Drake[20], F. Ebrahimi[21], J. Egedal[22], R. Ergun[6], G. Eyink[23], Y. Fan[24], G. Fiksel[25], C. Forest[22], W. Fox[21], D. Froula[8], K. Fujimoto[26], L. Gao[19], K. Genestreti[27], S. Gibson[24], M. Goldstein[28], F. Guo[17], J. Hare[29], M. Hesse[30], M. Hoshino[31], Q. Hu[32], Y.-M. Huang[1], J. Jara-Almonte[21], H. Karimabadi[33], J. Klimchuk[2], M. Kunz[1], K. Kusano[34], A. Lazarian[22], A. Le[17], S. Lebedev[35], H. Li[17], X. Li[36], Y. Lin[37], M. Linton[7], Y.-H. Liu[36], W. Liu[38], D. Longcope[39], N. Loureiro[29], Q.-M. Lu[40], Z-W. Ma[41], W. H. Matthaeus[42], D. Meyerhofer[17], F. Mozer[4], T. Munsat[6], N. A. Murphy[18], P. Nilson[8], Y. Ono[31], M. Opher[43], H. Park[44], S. Parker[6], M. Petropoulou[1], T. Phan[4], S. Prager[1], M. Rempel[24], C. Ren[8], Y. Ren[21], R. Rosner[45], V. Roytershteyn[18], J. Sarff[22], A. Savcheva[18], D. Schaffner[46], K. Schoeffier[47], E. Scime[12], M. Shay[42], L. Sironi[14], M. Sitnov[48], A. Stanier[17], M. Swisdak[20], J. TenBarge[1], T. Tharp[49], D. Uzdensky[6], A. Vaivads[50], M. Velli[51], E. Vishniac[23], H. Wang[13], G. Werner[6], C. Xiao[52], M. Yamada[21], T. Yokoyama[31], J. Yoo[21], S. Zenitani[53], E. Zweibel[22]

*[1]Princeton U., [2]NASA GSFC, [3]U. Iowa, [4]UC Berkeley, [5]Caltech, [6]U. Colorado – Boulder, [7]Naval Research Laboratory, [8]U. Rochester, [9]Swarthmore College, [10]Max Planck for Solar System Research, Germany, [11]Southwest Research Institute, [12]West Virginia U., [13]New Jersey Inst. Tech, [14]Columbia U., [15]Wheaton College, [16]Universities Space Research Association (USRA), [17]Los Alamos National Laboratory, [18]Harvard Smithsonian Center for Astrophysics, [18]Space Science Institute, [20]U. Maryland, [21]Princeton Plasma Physics Laboratory, [22]U. Wisconsin – Madison, [23]Johns Hopkins U., [24]High Altitude Observatory, [25]U. Michigan, [26]Beihang U., China, [27]U. New Hampshire, [28]Univ. Maryland Balt. County, [29]MIT, [30]U. Bergen, Norway, [31]U. Tokyo, Japan, [32]U. Alabama – Huntsville, [33]Analytics Ventures, [34]Nagoya U., Japan, [35]Imperial College, UK, [36]Dartmouth College, [37]Auburn U., [38]Lockheed Martin Solar & Astro Lab (LMSAL), [39]Montana State U., [40]U. Sci. Tech. of China, [41]Zhejiang U., China, [42]U. Delaware, [43]Boston U., [44]Ulsan Nat. Inst. Sci. Tech., Korea, [45]U Chicago, [46]Bryn Mawr College, [47]Instituto Superior Technico, Portugal, [48]APL, [49]Marquette U., [50]Swedish Institute of Space Physics, [51]UCLA, [52]Peking U., China, [53]Kobe U., Japan*

(108 authors and 53 institutions)

# Major Scientific Challenges and Opportunities in Understanding Magnetic Reconnection and Related Explosive Phenomena in Solar and Heliospheric Plasmas

## I. Magnetic Reconnection: A Fundamental Process throughout the Universe and in the Lab

Magnetic reconnection – the topological rearrangement of magnetic fields – underlies many explosive phenomena across a wide range of natural and laboratory plasmas [1-3]. It plays a pivotal role in electron and ion heating, particle acceleration to high energies, energy transport, and self-organization. Reconnection can have a complex relationship with turbulence at both large and small scales, leading to various effects which are only beginning to be understood. In heliophysics, magnetic reconnection plays a key role in solar flares, coronal mass ejections and heating, the interaction of the solar wind with planetary magnetospheres, associated dynamical phenomena such as magnetic substorms, and the behavior of the heliospheric boundary with the interstellar medium. Magnetic reconnection is also integral to the solar and planetary dynamo processes. In short, magnetic reconnection plays a key role in many energetic phenomena throughout the Universe, including extreme space weather events that have significant societal impact and laboratory fusion plasmas intended to generate carbon-free energy.

## II. Major Scientific Challenges in Understanding Reconnection and Related Explosive Phenomena in Heliophysics

**1. The multiple scale problem** [e.g., 3-19]: Reconnection involves the coupling between the fluid or MHD scale of the system and the kinetic ion and electron dissipation scales that are orders of magnitude smaller. This coupling is currently not well understood, and the lack of proper treatments in a self-consistent model is the core of the problem. Reconnection phase diagrams [e.g., 3] based on plasmoid dynamics clarify different possibilities for coupling mechanisms. Key questions include: how do plasmoid dynamics scale with key parameters, such as the Lundquist number and effective size; how is this scaling influenced by a guide field; do other coupling mechanisms exist; and how does reconnection respond to turbulence and associated dissipation on scales below or above the electron scales?
**2. The 3D problem** [e.g., 15-23]: Numerous studies have focused on reconnection in 2D while natural plasmas are 3D. It is critical to understand which features of 2D systems carry over to 3D, and which are fundamentally altered. Effects that require topological analysis include instabilities due to variations in the third direction leading to complex interacting "flux ropes", potentially enhancing magnetic stochasticity, and field-line separation in 3D. How fast reconnection is related to self-organization phenomena such as Taylor relaxation, as well as the accumulation of magnetic helicity, remains a longstanding problem with important implications for, e.g., coronal heating and eruptions.
**3. Energy conversion** [e.g., 24-35]: Reconnection is invoked to explain the observed conversion of magnetic energy to heat, flow, and to non-thermal particle energy. A major challenge in connecting theories and experiments to observations is the ability to quantify the detailed energy conversion and partitioning processes. Competing theories of particle acceleration based on 2D and 3D reconnection have been proposed, but as of yet there is no consensus on the origin of the observed power laws in particle energy distributions.
**4. Boundary conditions** [e.g., 36-39]: It is unclear whether an understanding of reconnection physics in periodic systems can be directly applied to natural plasmas, which are non-periodic and often line-tied at their ends such as in solar flares. Whether line-tying and driving from the boundaries fundamentally alter reconnection physics has profound importance in connecting laboratory physics to heliophysics. It is also important to learn how reconnection works in naturally occurring settings that have background flows, out-of-plane magnetic fields, and asymmetries.
**5. Onset** [e.g., 40-45]: Reconnection in heliophysical and laboratory plasmas often occurs impulsively, with slow energy build up followed by a rapid energy release. Is the onset a local, spontaneous (e.g., plasmoid instability) or a globally driven process (e.g., ideal MHD instabilities), and is the onset mechanism a 2D or 3D phenomenon? How do collisionality and global magnetic geometry affect the onset conditions? A related question is how magnetic energy is accumulated and stored prior to onset, e.g., in filament channels on the Sun and in the lobes of Earth's magnetotail.
**6. Partial ionization** [e.g., 46-51]: Reconnection events often occur in weakly ionized plasmas, such as the solar chromosphere (whose heating requirements dwarf those of corona), introducing new physics from neutral particles. Questions include whether reconnection is slowed by increased friction or accelerated by enhanced two-fluid effects.
**7. Flow-driven** [e.g., 44,52,53]: Magnetic fields are generated by dynamos in flow-driven systems such as stars and planets, and reconnection is an integral part of the dynamo process. Key questions include: under what conditions can reconnection occur in such systems; how fast does it proceed; how does reconnection affect the associated turbulence?
**8. Turbulence, shocks, and reconnection** [e.g., 14-17,54-61] Reconnection is closely interconnected to other fundamental plasma processes such as turbulence and shocks, which in turn produce heliophysical phenomena such as solar energetic particles. It is essential to understand the rates of topology change, energy release, and heating during reconnection, as they may be tied to the overall turbulence and shock dynamics.

**9. Related explosive phenomena** [e.g., 36,41,62] Global ideal MHD instabilities, both linear (kink, torus) and nonlinear (ballooning), are closely related to reconnection either as a driver or a consequence (e.g., coronal mass ejections, magnetic storms/substorms, and dipolarization fronts in the magnetotail). Understanding how, and under what conditions, such explosive phenomena take place, as well as their impact, remain major scientific challenges. Physics insights from reconnection under extreme conditions in astrophysics [e.g., 63] should be beneficial as well.

## III. Major Research Opportunities to Solve Challenges towards Helio2030, 2040 and 2050 Visions

| | Observations | Lab experiments | Computational and data science | Workforce/Diversity/Inclusion/Equity |
|---|---|---|---|---|
| Opportunities to Solve Major Scientific Challenges | Provide ground-truth of reconnection-relevant heliophysical phenomena but with limited access, control, and diagnostics. | Provide well-diagnosed and well-controlled data from real plasmas but limited in parameter space, geometry, and boundary conditions. | Provide flexibilities in simulating and analyzing underlying physics and databases, but in limited ranges of parameters with necessary assumptions. | Faster progress from a diversity of voices, societal benefits from a workforce that better reflects the US population. |
| 2030 Vision | Select new solar-eruption imaging mission with high resolution (0.1”) /cadence (~1s) HXR and context imaging, matching or exceeding DKIST, and in synergy with other ground-based observatories: GST, EOVSA, FASR, ngVLA. | Establish joint programs with DoE/NSF to support and use frontier lab facilities such as FLARE etc. to attack solar and heliospheric reconnection problems in wide ranges of collisionality, effective plasma size, and advanced diagnostics such as kinetic measurements. | Develop new models to capture the multiscale and 3D nature of reconnection phenomena, by strengthening collaborations with computer science and by expanding NASA high-performance computing capabilities and access to DoE, DoD, and NSF computing facilities. | Establish strong community support and NASA programs, such as CubeSat programs in which workforce development/diversity/equity/inclusion are essential criteria for success. |
| | Plan and select new geospace/solar wind constellation missions capable of measuring multiscale reconnection phenomena. | Explore and develop laboratory capabilities of modeling important heliophysical reconnection phenomena, such as CMEs, solar jets and global planetary magnetospheric reconnection. | Establish support and accessibility for databases and software repositories, utilize data science tools (AI/ML) for multi-scale reconnection research, and develop data-constrained/driven simulation techniques as vital community resources. | Increase positions at universities, HBCUs, etc. to attract and instruct next-generation researchers, especially theorists, model developers, and experimentalists. |
| 2040 Vision | Solar/solar wind multi-spacecraft missions in operation to provide critical data, including full-Sun high-resolution imaging and coronal magnetic field, with transmission of large datasets enabled. | Form an evolving network of frontier laboratory experiments with advanced diagnostics capabilities covering wide ranges of parameters, geometries and boundary conditions relevant to heliophysical reconnection phenomena. | Lead time and accuracy of space weather predictions improve substantially due to better understanding of reconnection, and routine incorporation of data driving and machine learning techniques. | Growing community participation with diversity/equity/inclusion, supported by NASA missions and programs. |
| 2050 Vision | Understanding and real-time prediction of space weather events by an effective interdisciplinary network of reconnection experts and other researchers, based on well-resolved comprehensive observations, simulations, lab experiments, and data assimilation, from a united, diversified, and well-balanced strong community. | | | |

In summary, magnetic reconnection is a fundamental process both in the Heliosphere and in laboratory plasmas. New research capabilities in theory/simulations, observations, and laboratory experiments will allow us to solve the grand scientific challenges summarized in this white paper. Success will require enhanced and sustained investments from relevant funding agencies, increased interagency/international partnerships, expanding/diversifying the solar, heliospheric, and laboratory plasma communities, as well as close collaborations between them. These investments will deliver transformative progress in understanding magnetic reconnection and related explosive phenomena in heliophysics, which will benefit many areas of critical practical importance, such as space weather.